\documentclass[
    aps,
    prl,
    reprint,
    superscriptaddress,
    amsmath,
    amssymb,
    longbibliography
]{revtex4-2}

\usepackage{graphicx}
\usepackage{dcolumn}
\usepackage{bm}
\usepackage{xcolor}
\usepackage{hyperref}
\usepackage{mathrsfs}
\usepackage{cancel}
\usepackage{array, makecell} %
\usepackage{comment}
\usepackage{tikz-cd}
\usepackage{hhline}
\usepackage{comment}
\usepackage{mathrsfs}
\usepackage{enumitem}
\usepackage{empheq}
\usepackage[all]{xy}
\usepackage{stmaryrd}
\usepackage{rotating}
\usepackage{color}  
\usepackage{slashed}

\newcommand{\p}{\partial}
\newcommand{\lbr}{\llbracket}
\newcommand{\rbr}{\rrbracket}

\newcommand{\cQ}{{\cal Q}}

\newcommand{\cD}{{\mathcal D}}

\newcommand{\cO}{{\mathcal O}}

\newcommand{\be}{\begin{equation}}
\newcommand{\ee}{\end{equation}}
\newcommand{\bs}{\begin{subequations}}
\newcommand{\es}{\end{subequations}}

\newcommand{\la}{\label}
\def\bea#1\eea{\begin{align}#1\end{align}}
\newcommand{\f}{\frac}
\newcommand{\bz}{{\bar z}}
\newcommand{\scri}{{\mathscr I}}

\newcommand{\QL}{Q^{\Lambda}_{-2}}
\newcommand{\GL}[1]{G^-_{\Lambda,#1}}
\newcommand{\GS}[1]{\mathcal{G}^-_{\Lambda,#1}}

\newcommand{\binomg}[2]{\binom{#1}{#2}}

\newcommand{\db}{\bar\partial}

\newcommand{\GTZ}[1]{G^{-,\mathrm{TZ}}_{#1}}

\begin{document}

\title{Higher-spin charges \& the $\mathcal{L}_{\Lambda} w_{1 + \infty}$ algebra in (A)dS$_4$}

\author{Lorenzo Di Giacomo}
\email{digiacomo.lorenzo@spes.uniud.it}
\affiliation{
Universit\`a degli Studi di Udine, via Palladio 8, I-33100 Udine, Italy
}
\affiliation{Institute for Fundamental Physics of the Universe (IFPU), Via Beirut 2, 34151 Trieste, Italy}
\affiliation{Istituto Nazionale di Fisica Nucleare, sezione di Trieste, Italy}

\author{Mait\'a Micol}
\email{maita.micol@kcl.ac.uk}
\affiliation{
Department of Mathematics, King's College London,
Strand, London, WC2R 2LS, United Kingdom
}

\author{Daniele Pranzetti}
\email{daniele.pranzetti@uniud.it}
\affiliation{
Universit\`a degli Studi di Udine, via Palladio 8, I-33100 Udine, Italy
}
\affiliation{Institute for Fundamental Physics of the Universe (IFPU), Via Beirut 2, 34151 Trieste, Italy}

\author{Ana-Maria Raclariu}
\email{ana-maria.raclariu@kcl.ac.uk}
\affiliation{
Department of Mathematics, King's College London,
Strand, London, WC2R 2LS, United Kingdom
}

\date{\today}

\begin{abstract}
We consider (3+1)-dimensional asymptotically locally (anti-)de Sitter ((A)dS$_4$) spacetimes with boundary conditions allowing for gravitational flux. We construct an infinite tower of higher-spin charges as perturbative solutions in the cosmological constant $\Lambda$ to a hierarchy of evolution equations resulting from an asymptotic expansion of the Einstein equations. 
We show that these charges canonically realize the $\Lambda$-deformed $w_{1+\infty}$ algebra ($\mathcal{L}_{\Lambda} w_{1+\infty}$) on a restricted gravitational phase space, thereby extending the higher-spin symmetry structure of asymptotically flat spacetimes to asymptotically (A)dS spacetimes. We further construct the curved-space counterparts of conformally soft gravitons directly in terms of spacetime data. We show that {they are primaries with respect to an $\mathfrak{sl}(2,\mathbb{R})$ subalgebra of the (A)dS$_4$ isometry algebra} and that their Poisson brackets with the quadratic higher-spin charges reproduce the $\Lambda$-deformed celestial operator product expansion proposed by Taylor and Zhu. Our results establish the bulk gravitational origin of the $\mathcal{L}_{\Lambda}w_{1+\infty}$ symmetry and extend the celestial bulk-boundary dictionary to asymptotically (A)dS$_4$ gravity.
\end{abstract}

\maketitle

\textit{Introduction.}--- 
Symmetries are among the key data characterizing physical theories, with far-reaching implications. In quantum field theories, global symmetries constrain spectra and dynamics \cite{Wigner:1939cj,PhysRevLett.18.188,Belavin:1984vu,Coleman:1985rnk,Gaiotto:2014kfa}, while an abundance of symmetry can in some cases almost completely determine the theory \cite{Zamolodchikov:1978xm, Mason:1991rf}. In gravity, global symmetries are conjectured to be absent \cite{PhysRevD.83.084019,Harlow:2018tng}, but large-gauge or asymptotic symmetries may still exist \cite{Bondi:1960jsa,Bondi62,Sachs:1962wk,Penrose:1962ij,Brown:1986nw, Henneaux:1985tv, Barnich:2009se}. The latter correspond to the component of the diffeomorphism group that survives upon gauge fixing and are often closely tied to the existence of boundaries and choice of boundary conditions. Large-gauge symmetries generally imply the existence of physical charges which are part of the observable data of a theory \cite{Ashtekar:1981bq, Iyer:1994ys,Wald:1999wa, Barnich:2001jy}.  Within the holographic paradigm in which theories of gravity are expected to admit a dual description in terms of quantum field theories \cite{Susskind:1994vu,Maldacena:1997re,Aharony:1999ti}, the global symmetries of the latter are realized as large-gauge symmetries of the former. 

This phenomenon has been studied extensively in $(3+1)$-dimensional (4d) asymptotically flat spacetimes (AFS). In this context, the asymptotic symmetries consist of Poincar\'e symmetries and an infinite-dimensional, local enhancement thereof including BMS$_4$ symmetries \cite{Bondi:1960jsa,Bondi62,Sachs:1962wk}. The leading and subleading soft graviton theorems revealed that these symmetries are dual to chiral algebras in 2d conformal field theory (CFT$_2$), notably including Virasoro \cite{Strominger:2013jfa,He:2014laa,Kapec:2014opa, Kapec:2016jld}. More recently, a whole tower of subleading soft graviton modes was found to yield a further enhancement to an infinite higher-spin symmetry algebra, namely the (loop algebra of the wedge subalgebra of the) $w_{1 + \infty}$ algebra ($\mathcal{L}w_{1+\infty}$) \cite{Guevara:2021abz, Strominger:2021mtt}. A spacetime understanding of this algebra has been developed from both perturbative/canonical gravity \cite{Freidel:2021dfs, Freidel:2021ytz,Geiller:2024bgf, Cresto:2024fhd, Cresto:2024mne} and twistor space  \cite{Adamo:2021lrv,Kmec:2024nmu,Kmec:2026dis} perspectives. In the bulk, the higher-spin symmetries are realized by spacetime charges constructed as light-ray operators involving the radiative data at future (past) null infinity $\mathscr{I}^+$  ($\mathscr{I}^-$). 

These rich symmetry structures are often associated with the presence of a null boundary, and not known to arise in 4d asymptotically constant curvature backgrounds, such as (anti)-de Sitter spacetimes ((A)dS$_4$), at least with the standard choice of Dirichlet boundary conditions. Nevertheless, building on \cite{Compere:2008us, Poole:2018koa}, it was shown in \cite{Compere:2019bua, Compere:2020lrt} that relaxing the boundary conditions to allow for flux through a finite interval of the conformal boundary does lead to an infinite-dimensional local enhancement of the (A)dS$_4$ isometry algebra, now known as the $\Lambda$-BMS algebra. This algebra was shown to reduce to the extended BMS algebra of 4d AFS in the limit of vanishing cosmological constant $(\Lambda = 0)$. The goal of this work is to establish whether the $\Lambda$-BMS algebra may be further enlarged by a counterpart of the higher-spin charges previously constructed in AFS \cite{Freidel:2021dfs, Freidel:2021ytz}.

The motivation to expect a positive answer comes from the fact that the $w_{1 + \infty}$ algebra admits several deformations \cite{Pope:1989ew,Fairlie:1990wv}. Relevant for this work will be its $\Lambda$-deformation, the $\mathcal{L}_{\Lambda}w_{1 + \infty}$ algebra, which was found in \cite{Taylor:2023ajd} to arise from a rather simple, $\mathfrak{sl}(2,\mathbb{R})$ covariant modification of the OPE of conformally soft gravitons in 4d flat spacetime. 
From a twistor perspective, it was shown in \cite{Bittleston:2024rqe} that a non-vanishing cosmological constant yields the same deformation of the $w_{1 + \infty}$ algebra. More recently, an intrinsically  CFT$_3$ realization in terms of light-ray operators associated with the stress tensor was proposed in \cite{Strominger:2026yrh}. However, a bulk understanding of this algebra in asymptotically locally (A)dS$_4$ spacetimes, including the choice of boundary conditions and gravitational phase space that would allow for an explicit, first principles construction of the associated higher-spin charges, remained open. In this work we will fill this gap. 

We now outline our strategy and main results. We consider solutions to the Einstein equations (EE) with a cosmological constant $\Lambda \neq 0$. We will be interested in their asymptotic expansion near the conformal boundary. It will be convenient to work in the Bondi gauge \cite{Bondi:1960jsa} in which the metric takes the general form \eqref{BSLineElement}. The nature of the conformal boundary at $r \rightarrow \infty$ changes depending on the sign of $\Lambda$: this is timelike for $\Lambda < 0$ and spacelike for $\Lambda > 0$. The retarded time $u$ labels outgoing null hypersurfaces and becomes either a time coordinate on the AdS boundary for $\Lambda < 0$ or a spatial coordinate at the future boundary of dS for $\Lambda > 0$. Following \cite{Compere:2019bua}, we will always denote the conformal boundary by $\mathscr{I}$. The sign of $\Lambda$ will not play any role beyond these physical considerations. Employing the Newman-Penrose formalism \cite{Newman:1961qr,Newman:1968uj}, we first show that the asymptotic EE at leading order in the $r^{-1}$ expansion can be recast as the recursion relations \eqref{Lrec} for $s \leq 2$. Remarkably, the effect of curvature is encoded in a simple $\Lambda$-correction to the expansion of the EE near $\mathscr{I}$ in AFS \cite{Barnich:2019vzx,Freidel:2021qpz, Freidel:2021ytz}. 

Inspired by the construction of the higher-spin charges of \cite{Freidel:2021ytz}, we extend these equations to all integer $s \geq 0$ and study their solutions perturbatively, to linear order in $\Lambda$. Just as in AFS, the resulting charges diverge as $|u| \rightarrow \infty$. To construct finite charges, we employ the procedure developed in \cite{Cresto:2024fhd, Cresto:2024mne} and define a master charge \eqref{cQ} required to be conserved in a non-radiative interval on $\mathscr{I}$. Conservation imposes a hierarchy of dual recursion relations \eqref{puT} on the parameters $T_n$, subject to which the master charge reduces to 
\eqref{QL0}. Our main result is that these charges provide a representation of $\mathcal{L}_{\Lambda} w_{1 + \infty}$ on the (A)dS$_4$ phase space restricted to the radiative data encoded in the negative helicity component of boundary stress tensor and its canonically conjugate component of the boundary metric.  

The proof contains three main parts. We first introduce the $\Lambda$-deformed bracket \eqref{Lbra} for the dual parameters and demonstrate that the action of the master charge on the boundary data provides a representation of this bracket when the dual recursion relations \eqref{puT} are obeyed. We then use the solutions to the dual recursion relations to explicitly construct the $\Lambda$-deformed higher spin charges at linear order in $\Lambda$ and quadratic order in the phase space data and show that they satisfy the $\mathcal{L}_{\Lambda} w_{1 + \infty}$ algebra. These results generalize those of \cite{Cresto:2024fhd, Cresto:2024mne} and \cite{Freidel:2021dfs, Freidel:2021ytz} to asymptotically maximally symmetric spacetimes. Finally, we introduce a novel $\Lambda$-deformation of the flat space conformally soft gravitons \cite{Guevara:2019ypd,Pate:2019lpp,Puhm:2019zbl, Freidel:2021ytz}, which transform as primaries with respect to an $\mathfrak{sl}(2,\mathbb{R})$ subalgebra of the (A)dS$_4$ isometry algebra. We then 
show that their Poisson bracket with the $\Lambda$-deformed quadratic (or hard) charges reproduce precisely the $\Lambda$-deformed OPE proposed by Taylor and Zhu in \cite{Taylor:2023ajd}. In the remainder of this letter, we further explain our approach and present our main results. All detailed computations will appear in an accompanying paper \cite{DiGiacomoEtAl2026}. 

\textit{Bondi gauge and restrictions on the phase space.}--- 
We consider asymptotically locally (A)dS$_4$ \cite{deHaro:2000vlm,Poole:2018koa,Compere:2020lrt,Compere:2019bua} spacetimes in Bondi--Sachs (BS) coordinates $(u,r,x^A)$. 
After imposing the gauge conditions
\begin{equation}\label{BSgauge}
    g_{rr}=0,\qquad g_{rA}=0,\qquad\partial_r\left(\frac{\text{det}g_{AB}}{r^4}\right)=0,
\end{equation}
the line element takes the general form
\begin{equation}\label{BSLineElement}
    \begin{split}
        ds^2&=e^{2\beta}\frac Vr d u^2 - 2e^{2\beta} du dr\\ &+g_{AB}\big(dx^A-U^Adu\big)\big(dx^B-U^Bdu\big).
    \end{split}
\end{equation}
We are interested in asymptotic solutions to the Einstein equations. Following \cite{Compere:2019bua}, we consider the large-$r$ expansion near the conformal boundary $\mathscr{I}$ \footnote{Strictly speaking, $\mathscr{I}$ is the portion of conformal boundary reached by the null rays identified by $(u,x^A)$. For $\Lambda<0$ ($\Lambda>0$) it is a timelike (spacelike) surface of codimension 1.}
\begin{equation}\label{metricexpanison}  g_{AB}=r^2q_{AB}+rC_{AB}+D_{AB}+r^{-1}E_{AB}+\mathcal{O}(r^{-2})
\end{equation} and first solve the radial equations $G_{r\alpha}+6\Lambda g_{r\alpha}=0$, where $6\Lambda \equiv \pm \frac{3}{\ell^2}$ is the cosmological constant and $\ell$ is the (A)dS radius. These equations determine the large-$r$ behaviour of the functions appearing in \eqref{BSLineElement}.
It was shown in \cite{Compere:2019bua} that the leading coefficients of $\beta$ and $U^A$ remain unconstrained and can therefore be chosen to vanish. 
As a result, the functions parametrizing the metric obey the fall-offs
\begin{equation}
\begin{split}
    \beta &\sim \mathcal{O}(r^{-1}), \quad U^A \sim \mathcal{O}(r^{-1}), \quad \frac{V}{r} \sim  \mathcal{O}(r^2).
    \end{split}
\end{equation}
On the other hand, the equations $G_{AB}+6\Lambda g_{AB}=0$ provide the evolution along $u$ of the coefficients in \eqref{metricexpanison}. In \cite{Compere:2019bua} it was shown that the data $\{q_{AB},E^{\small{\text{TF}}}_{AB}\}$ remain unconstrained for $\Lambda\neq0$, where $E^{\small{\text{TF}}}_{AB}$ is the trace-free part $E_{AB}$~\footnote{The degrees of freedom in $E^{TF}_{AB}$ can be expressed via the auxiliary quantity $J_{AB}$, the traceless, angular part of the holographic stress energy tensor. Then, $\{q_{AB},J^{CD}\}$ form a canonically conjugated pair. Please refer to \cite{Compere:2019bua} for more details.}. Moreover, $C_{AB}$ is identified with the trace--free part of $\partial_uq_{AB}$. We will further set 
\begin{equation}\label{pureshearcondition}
   q^{AB}\partial_uq_{AB}=0, 
\end{equation}
which implies the key relation
\be
 \partial_uq_{AB}=2\Lambda C_{AB}.
 \la{puq}
\ee

Throughout this letter, we will use the Newman-Penrose (NP) formalism \cite{Newman:1968uj}. This allows for the non-radial EE to be recast as evolution equations for a set of Weyl scalars, defined by projecting the Weyl tensor onto components of a null frame. We denote by $m_0^A, \bar m_0^A$  the leading NP null dyad, satisfying $q^{AB}=m_0^{A}\bar m_0^{B}+\bar m_0^{A} m_0^{B}$. The dyad
is defined up to local $U(1)$ rotations $m_0 \rightarrow e^{i \theta}m_0,~ \bar{m}_0 \rightarrow e^{-i \theta}\bar{m}_0$.
We use this freedom, together with \eqref{pureshearcondition}, to set the leading spin coefficient
\begin{equation}
    \gamma_0=0.
\end{equation}
Following \cite{Freidel:2021ytz}, we complexify the phase space and restrict to the single-helicity sector
\begin{equation}\label{singlehelicity}
    \bar{C}=C_{AB}\bar{m}_0^A\bar{m}_0^B=0,
\end{equation}
in which case the dyad satisfies the evolution equations
\begin{equation}\label{dyadevolution}
    \partial_um_{0A}=\Lambda C\bar{m}_{0A},\qquad\partial_u\bar{m}_{0A}=0,
\end{equation}
where $C=C_{AB}m_0^Am_0^B$. In the restricted phase space defined by  \eqref{pureshearcondition}, \eqref{singlehelicity} and neglecting $\mathcal{O}(\Lambda^2)$ terms, the degrees of freedom $\{q_{AB}, E^{\small{\text{TF}}}_{AB}\}$ reduce to the canonical pair $\{\partial_u^{-1}C,Q_{-2}\}$, where $Q_{-2}$ is identified with the leading order coefficient of the Weyl scalar $\Psi_4$ and we introduced the notation  $\p_u^{-1}f(u):=\int_{+\infty}^udu' f(u')$. 

\textit{Recursive  equations for charge evolution.}--- 
We would like to investigate if the construction of higher-spin charges in AFS \cite{Freidel:2021ytz} extends to asymptotically locally (A)dS$_4$ spacetimes. 
Considering the large-$r$ expansion of the NP scalars
\begin{equation}
    \Psi_i = \sum_{k = 0}^{\infty} \frac{\Psi_i^{(k)}}{r^{5 - i + k}}, \quad 0\leq i \leq 4,
\end{equation}
 we define the covariant functionals or charge aspects $ Q _s=\Psi^{(0)}_{2-s}$,
 for $-2 \leq s\leq 2$. The $u$-evolution Bianchi identities yield the recursion relations \cite{Geiller:2022vto}
 \begin{equation}\label{recEq0}
    \p_u Q _s=\eth Q _{s-1} +\frac{(s+1)}{2}C Q _{s-2}+\Lambda \Psi^{(1)}_{2-s},
\end{equation}
where $s$ is the helicity \footnote{In going from spin-$j$ tensors to helicity-$s$ scalars through the contraction $O_s=O_{A_1\dots A_j} m_0^{A_1}\dots m_0^{A_j}$, with $s(m_0^A)=+1$, the first label is traded for the second; negative helicity-$s$ scalars are obtained through the contraction with the complex conjugate dyad as $s(\bar m_0^A)=-1$.} and
\begin{align}
    &\eth = 
    m_0^A\partial_A-s D_Am_0^A. 
    \la{eth}
\end{align}
 Moreover, \eqref{dyadevolution} implies that $\eth$ does not commute with $\p_u$ but instead satisfies 
 \begin{equation}
    [\p_u,\eth]\eta_s=\Lambda(- C\bar\eth+s\bar\eth C)\eta_s,
    \la{comm}
\end{equation}
when acting on a helicity-$s$ field $\eta_s$.

The sub-leading Weyl scalar  terms appearing on the RHS of \eqref{recEq0} were derived in \cite{Saw:2016isu,Mao:2019ahc} and allow for the charge evolution equations \eqref{recEq0} to be rewritten as
\be
\p_u Q _s=\eth Q _{s-1}+\frac{(s+1)}2 C Q _{s-2}-\Lambda\bar \eth Q_{s+1},
\la{Lrec}
\ee
after imposing \eqref{singlehelicity} and identifying $\Psi_0^{(1)}=-\bar \eth{ Q}_{3}$. For $s=0,1$ one recovers the evolution equations for the mass and the angular momentum obtained in \cite{Compere:2019bua} in vectorial form. The charge aspect $Q_{-2}$ encodes radiative free data. 

In the following, we investigate the implications of extending the validity of \eqref{Lrec}
for all integer $s\geq-1$. In the spirit of \cite{Freidel:2021ytz}, we expect the evolution equations \eqref{Lrec} for $s > 2$ to capture subleading orders in $r^{-1}$ of the Bianchi identities for $\Psi_0$ with
\be
\Psi_0^{(k)}=\f{(-1)^k}{k!}{\bar\eth}^{k} { Q}_{k+2} +\cdots
\ee
and upon truncating to terms linear in $C$ and/or $\Psi$. For $s=3$, one can verify that \eqref{Lrec} is exact modulo the restriction $\bar C = 0$,
which would otherwise introduce an extra cubic term proportional to $\Lambda$ \cite{DiGiacomoEtAl2026}.
For $\Lambda \neq 0$, \eqref{Lrec} determines $\bar{\eth} Q_{s+1}$ algebraically in terms of the lower-spin data. On the other hand, for $\Lambda = 0$ they reduce to the hierarchy of equations whose solutions were shown to generate the $w_{1 + \infty}$ algebra of AFS \cite{Guevara:2021abz, Strominger:2021mtt}. To make contact to these works, it is hence natural to instead search for perturbative solutions in $\Lambda$ to \eqref{Lrec}. To linear order in $\Lambda$, the $\Lambda \bar{\eth}Q_{s+1}$ term acts as a source entirely determined by the $\Lambda = 0$ data. One can then proceed to find the $\Lambda$-corrections to the tower of charges constructed in \cite{Freidel:2021ytz}. We will demonstrate that, after suitable renormalization, these charges generate the $\mathcal{L}_{\Lambda}w_{1 + \infty}$ algebra \cite{Taylor:2023ajd} on the aforementioned restricted phase space of asymptotically (A)dS$_4$ spacetimes. 


\textit{Master charge and dual EOM.}---
Just as in flat space \cite{Freidel:2021ytz}, all higher-spin charges determined by \eqref{Lrec} suffer from large-$u$ divergences. To construct  finite charges, we employ the systematic procedure developed in \cite{Cresto:2024mne}. 
We start by introducing the {\it master charge}
\be
 \cQ^\Lambda[T](u):=\f{8}{\kappa^2} \sum_{n=-1}^\infty \int_S T_n(u,z,\bz) Q_n(u,z,\bz),
 \la{cQ}
\ee
where $T_n(u,z,\bz)$ are dual parameters on $\scri$ 
of helicity $-n$ and $\kappa^2=32\pi G$.
The charges \eqref{cQ} are conserved on non-radiative cuts, defined by the condition $Q_{-2}=0$, provided that $Q_n$ satisfy \eqref{Lrec} and $T_n$ obey dual recursion relations 
\be
    \partial_uT_n =\eth T_{n+1}
 -\frac{(n+3)}{2}CT_{n+2}-\Lambda\bar \eth T_{n-1},\qquad n\geq-1,
 \la{puT}
\ee
with $T_{-2}=0$. As a result, the master charge on a generic $u$-cut of $\mathscr{I}$ can be expressed as
\be
 \cQ^\Lambda[T](u)=\f{8}{\kappa^2} \p_u^{-1} \int_S\left( T_{-1} \eth Q_{-2}+\f12T_{0}C Q_{-2}\right),
 \la{QL0}
 \ee
 where we imposed the boundary condition $\cQ^\Lambda[T](+\infty)=0$.
The symplectic structure of asymptotically locally (A)dS$_4$ spacetimes was discussed in \cite{Papadimitriou:2005ii, Compere:2008us,Poole:2018koa,Compere:2019bua,Compere:2020lrt}. In the helicity scalar formalism and subject to the phase space restrictions introduced before, the symplectic form is given by
\be
\label{eq:sf}
\Omega=-\f{4}{\kappa^2} \int_\scri \delta Q_{-2}\wedge \delta(\p_u^{-1}C).
\ee
Therefore, the master charge \eqref{cQ} can be expressed as a Noether charge 
\be
\label{eq:nc}
\cQ^\Lambda[T]=\lim_{u\to-\infty} \cQ^\Lambda[T](u)=-\f{4}{\kappa^2} \int_{\scri} Q_{-2} \delta^\Lambda_T (\p_u^{-1} C),
\ee
with 
 \be
  \delta^\Lambda_T (\p_u^{-1} C)= -2\eth T_{-1}+ T_{0}C.
  \la{dTuC}
 \ee
It follows that the variation of the shear under \eqref{QL0} takes the form
 \bea
 \begin{split}
 \delta^\Lambda_T C&= 
 \Lambda (C \bar \eth T_{-1}
 -2 T_{-1} \bar \eth C)
\\
&
 -2\eth^2 T_0 +2 T_1 \eth C
 +3C\eth T_1
 -\f32 C^2T_2
+ T_{0}\p_u C,
 \la{dTC}
 \end{split}
 \eea
 where we used \eqref{comm} and \eqref{puT}. This formula generalizes the transformation of the shear obtained in \cite{Cresto:2024mne} to include the effect of curvature to leading order in $\Lambda$.


\textit{$\Lambda$-deformed symmetry algebroid.}---
 For $\Lambda = 0$, it was shown in \cite{Cresto:2024mne} that the transformation \eqref{dTC} provides a representation of the symmetry algebroid bracket
\be
\lbr T',T \rbr_s:= [T',T]^C_s+(\delta_T T')_s-(\delta_{T'}T)_s\,,\quad s\geq -1,
\la{dbra}
\ee
for $T_n$ obeying the $\Lambda=0$ dual recursion relations \eqref{puT}.
Here 
\be
[T',T]^C_s:=\sum_{n=0}^{s+1}(n+1)(T'_n(\cD T)_{s-n}-T_n(\cD T')_{s-n})
\la{Cbra}
\ee
is the Schouten--Nijenhuis graded Lie bracket \cite{Norris1997SchoutenNijenhuis} and
\be
(\cD T)_s:=\eth T_{s+1}-\f{(s+3)}2 CT_{s+2}.
\ee
For $C = 0$, $T$ in degree $p$ and $T'$ in degree $q$, the bracket \eqref{Cbra} coincides with the $w_{1 + \infty}$ bracket in degree $s= p + q-1$ \cite{Cresto:2024fhd}.

Motivated by the well-known $\Lambda$-deformation of $\mathcal{L}w_{1+\infty}$ \cite{Taylor:2023ajd,Bittleston:2024rqe} reviewed in the Appendix (equation \eqref{eq:charge-aspect-alg}), we introduce the $\Lambda$-deformed symmetry algebroid bracket for $s\geq -1$
\bea
\begin{split}
\lbr T',T \rbr^\Lambda_s&:= \lbr T',T \rbr_s 
\cr
&
+\Lambda \sum_{n=-1}^{s}(n-1)(T'_n\bar \eth  T_{s-1-n}-T_n\bar \eth T'_{s-1-n}),
\la{Lbra}
\end{split}
\eea
with $\delta_{T}T'\to \delta_{T}^{\Lambda}T'$ in \eqref{dbra}. 
The first main result of this letter is the proof that the  $\Lambda$-deformed action of $\delta_T^{\Lambda}$ on $\partial_u^{-1}C$ provides a representation of \eqref{Lbra} on the support of the dual evolution equations \eqref{puT}, namely 
\begin{equation}
[\delta^{\Lambda}_{T'},\delta^{\Lambda}_{T}](\partial_u^{-1}C)\hat{=}-\delta^{\Lambda}_{\lbr T',T\rbr ^{\Lambda}}(\partial_u^{-1}C).
    \la{brarep}
\end{equation}
We outline the main steps leading to the result \eqref{brarep} and defer the detailed derivation to our accompanying paper \cite{DiGiacomoEtAl2026}. 
One first writes
\begin{equation}
\begin{split}
 &[\delta_{T'}^{\Lambda},\delta_T^{\Lambda}](\partial_u^{-1}C)+\delta^{\Lambda}_{\lbr T',T\rbr }(\partial_u^{-1}C)\\
&=-2[\delta^{\Lambda}_{T'},\eth]T_{-1}+2[\delta^{\Lambda}_{T},\eth]T'_{-1}
\\
&+T_0\delta_{T'}^{\Lambda}C-T'_0\delta_{T}^{\Lambda}C
 \\
&
 -2\eth[T',T]^{\Lambda}_{-1}+[T',T]^{\Lambda}_0C,
\end{split}
\end{equation}
where $\lbr T',T \rbr^\Lambda_s=[T',T ]^\Lambda_s+(\delta^\Lambda_T T')_s-(\delta^\Lambda_{T'}T)_s$.
The first line on the RHS can be computed using 
\begin{equation}
\begin{split}
[\delta_T^\Lambda,\eth]\eta_s&=-\Lambda\delta_T^\Lambda(\partial_u^{-1}C)\bar{\eth}\eta_s+\Lambda{s}\bar{\eth}\left(\delta_T(\partial_u^{-1}C)\right)\eta_s,
    \\
[\delta^\Lambda_T,\bar{\eth}]\eta_s&=0,
\end{split}  
\la{commdD}
\end{equation} 
the second line is given by  \eqref{dTC} and the third line is obtained using \eqref{Lbra}, \eqref{Cbra}, whereby the field dependent terms $\delta_{T'}^{\Lambda}T$ cancel. These three contributions combine into terms proportional to the dual evolution equations \eqref{puT} for $T_{-1}, T_0$ and therefore establish the result \eqref{brarep} on-shell.


\textit{Higher-spin charges and the $\mathcal{L}_{\Lambda}w_{1+\infty}$ algebra.}--- For $\Lambda = 0$, it was shown in \cite{Cresto:2024mne} that the charges \eqref{eq:nc} provide a canonical realization of the $\mathcal{L}w_{1 + \infty}$ algebra on the gravitational phase space. Direct contact with the construction of the higher-spin charge aspects of \cite{Freidel:2021ytz} was made by solving the dual recursion relations \eqref{puT} with the boundary condition $T_{n+1} = 0,~ n \geq s.$ In this section we demonstrate that this approach generalizes to asymptotically weakly curved backgrounds. Working to linear order in $\Lambda$ and $C$ we construct a tower of higher-spin charges and  show that they provide a canonical representation of the $\mathcal{L}_{\Lambda}w_{1 + \infty}$ algebra. This is our second main result.

We start by introducing a double expansion in $(\Lambda,C)$ of the parameters $T_k=T_k^{(0,0)}+T_k^{(0,1)}+\Lambda T_k^{(1,0)}+\Lambda T_k^{(1,1)}$ and chose as initial data for the spin-$s$ master charge 
\begin{equation}
\label{eq:in-da}
\begin{split}
    T^{(0,0)}_s(0,z,\bar z)&=\tau(z,\bar z), \\ 
    T^{(0,0)}_k(0,z,\bar z)&=0,\ \  \forall k\neq s\\
    T^{(0,1)}_k(0,z,\bar z)&=T^{(1,0)}_k(0,z,\bar z)=T^{(1,1)}_k(0,z,\bar z)=0\ \forall k.
\end{split}
\end{equation}
We fix the metric on a given cut $u = u_0$ to be the one of the plane, such that $\eth\lvert_{u_0}=\partial$. The solution to \eqref{puT} with these initial conditions is presented in the Appendix (equations \eqref{T00}, \eqref{eq:Tquadratic}). 
The higher-spin charges can then be obtained by direct substitution of these solutions into \eqref{eq:nc}, \eqref{dTuC}. For the linearized (${\cal O}(C^0)$) charges, one recovers the flat space result \cite{Freidel:2021ytz}
\begin{equation}
\label{eq:l0}
    \cQ_s^{(0,0)}[\tau] = \frac{8}{\kappa^2} \int du \int_S \frac{u^{s+1}}{(s+1)!} \partial^{s+2}\tau Q_{-2} 
\end{equation}
and its $\Lambda$-correction
\begin{equation}
\label{eq:l1}
    \cQ_s^{(1,0)}[\tau] = -\frac{8}{\kappa^2} (s+2) \int du \int_S  \frac{u^{s+3}}{(s+3)!} \bar{\partial} \partial^{s+3}\tau Q_{-2}.
\end{equation}
The $\Lambda$-corrected quadratic charge is presented in \eqref{eq:quadrlc}, \eqref{eq:qch}.
The algebra of these charges can then be computed using the Poisson brackets implied by \eqref{eq:sf}. The result is that these  $\Lambda$-corrected higher-spin charges obey the  $\mathcal{L}_{\Lambda}w_{1 + \infty}$ algebra \eqref{eq:cbl} at the linear perturbative order. The complete derivation will appear in our companion paper \cite{DiGiacomoEtAl2026}.


\textit{Derivation of the $\Lambda$-corrected ``celestial'' graviton OPE.}--- In \cite{Taylor:2023ajd} Taylor and Zhu (TZ) proposed a $\Lambda$-corrected OPE of conformally soft gravitons that implies the $\Lambda$-deformation of the $\mathcal{L}w_{1 + \infty}$ algebra derived in \cite{Guevara:2021abz,Strominger:2021mtt}. The proposed OPE \eqref{OPETZ} was notably not the natural extension of the flat space result expected from the scattering of gravitons on a curved background \cite{Alday:2022uxp,Alday:2022xwz,Alday:2023jdk,Alday:2023mvu}: the OPE coefficient of the $\Lambda$-correction needed to reproduce the $\mathcal{L}_{\Lambda} w_{1 + \infty}$ algebra was found to differ from the amplitudes expectation by a dimension-dependent factor. 
Building on our previous analysis, in this section we present a first-principles derivation of the TZ OPE. This is our third main result.

The $\Lambda$-corrected linearized charge obtained by summing \eqref{eq:l0} and \eqref{eq:l1} can be written as the $\p^{s+2}$  descendant of the $\Lambda$-deformed phase space variable 
\begin{equation}
\QL:=\Bigl[1+\Lambda(\p_u^{-1})^2\p\bar \p(\Delta+1)\Bigr]Q_{-2}, \quad \Delta=u\p_u+1
\la{QL}
\end{equation}
namely
\begin{equation}
\cQ_{s,\mathrm{lin}}^\Lambda[\tau]
= \frac{(-1)^s}{(s+1)!}\int_S\tau \p^{s+2}\int_{-\infty}^{+\infty} du\,u^{s+1}  Q^\Lambda_{-2}.
\la{Qlin}
\end{equation}
The key observation is that 
\begin{equation}
\GS{1 - s}
:= \frac{(-1)^s}{(s+1)!}\int_{-\infty}^{+\infty}  du\, u^{1+s} \QL
\la{GS}
\end{equation}
 is an $\mathfrak{sl}(2,\mathbb{R})$ primary of $\mathfrak{so}(3,1)$ dimension $1 - s$ and spin $-2$, namely it transforms under the action of $\mathfrak{sl}(2,\mathbb{R}) \subset \mathfrak{so}(3,1) \subset \mathfrak{so}(3,2)$ as 
 \begin{equation}
 \begin{split}
     \delta_{L_1}\GS{1 - s} &=\left[-(s+1) z + z^2 \partial \right] \GS{1 - s}, \\
     \delta_{L_0}\GS{1 - s} &=\left[-\frac{1}{2}(s+1)  + z \partial \right] \GS{1 - s},\\
     \delta_{L_{-1}}\GS{1 - s} &= \p \GS{1 - s}.
     \end{split}
 \end{equation}
One can show this using the definitions \eqref{QL} and \eqref{GS} and the fact that $Q_{-2}$ is a 3d conformal primary of dimension $\Delta_0 = 3$ and spin $J = -2$ transforming as
 \begin{equation}
     \delta_{L_1} Q_{\Delta_0,J} = \left[(\Delta_0 + u \p_u + J) z + z^2 \p -\Lambda u^2\bar{\partial}\right] Q_{\Delta_0, J}.
 \end{equation}
 Consequently, \eqref{Qlin} is in fact an $\mathfrak{sl}(2, \mathbb{R})$ primary descendant \cite{Gelfand-Book,Pasterski:2021fjn}. Just as in flat space, \eqref{GS} can be obtained as the residue of a $\Lambda$-deformed conformal primary graviton
 \begin{equation}
\GL{\Delta}
:=-\Gamma(\Delta-2)\int_{-\infty}^{+\infty}  du\,(u+i\epsilon)^{2-\Delta}\QL
\la{GL}
\end{equation}
at integer $\Delta \leq 2$. The operators in \eqref{GS} are therefore the curved space counterparts of the flat space conformally soft gravitons \cite{Guevara:2019ypd,Pate:2019lpp,Puhm:2019zbl, Freidel:2021ytz} and provide a novel definition in terms of spacetime data of the celestial operators introduced by TZ. We conclude this section by demonstrating that the TZ OPE follows from this definition upon importing the flat space OPE-bracket correspondence \cite{Freidel:2021ytz,Pranzetti:2025flv} to this context.

In the presence of a cosmological constant, the OPE-bracket correspondence is expected to imply an equivalence between the transformation of the $\Lambda$-conformally soft graviton under the higher-spin symmetries discussed before
\begin{equation}
\label{eq:tr-cpg}
\delta_T^{\Lambda}\GL{\Delta_2}(z_2,\bz_2)=\{\cQ_{s}^\Lambda[\tau], \GL{\Delta_2}(z_2,\bz_2)\}
\end{equation}
and the celestial OPE
\bea
\int_{S}d^2z_1 \tau(z_1,\bz_1)\p^{s+2} \GS{1-s}(z_1,\bz_1)\GL{\Delta_2}(z_2,\bz_2).
\la{OPE/bra}
\eea
The Poisson bracket \eqref{eq:tr-cpg} can be computed using
\begin{equation}
\delta_T^\Lambda\QL
=\delta_T^\Lambda Q_{-2}
+\Lambda(\p_u^{-1})^2\p\bar \p(\Delta+1)\delta_T Q_{-2}
+\cO (\Lambda^2).
\la{dLQ}
\end{equation}
After a lengthy calculation which will be presented in our companion paper \cite{DiGiacomoEtAl2026}, we obtain 
\begin{align}
&\delta_T^{\Lambda}\GL{\Delta_2}
=\sum_{n=0}^{s}(n+1)\binomg{\Delta_2-2}{s-n}
(\p^{s-n}\tau)\p^n\GL{\Delta_2+1-s}\notag\\
&-\Lambda\sum_{n=0}^{s+1}\binomg{\Delta_2-3}{s+1-n}
\Bigl[(\Delta_2+1-s)(\p^{s+2-n}\db\tau)\p^n\notag\\
&+(1-s)(\p^{s+1-n}\tau)\p^n\db\Bigr]
\GL{\Delta_2-s-1}.
\la{dLG}
\end{align}
This can be shown to agree precisely with the TZ OPE of two 
 negative-helicity conformal primary gravitons \eqref{OPETZ}, after
taking the conformally soft limit on the first entry, applying $\int_{S} d^2 z_1 \tau(z_1,\bz_1)\p_{}^{s+2}$ and  
identifying $G_{\Delta}^{-,\mathrm{TZ}}
=i^\Delta\frac{8\pi}{i\kappa}\,G^-_{\Lambda,\Delta}$. This extends the bulk-boundary dictionary entry introduced in \cite{Freidel:2021ytz} in flat space to maximally symmetric curved backgrounds.


\textit{Discussion.}---
Our work opens up several new avenues to be explored in the future. First of all, it suggests that our understanding of the AdS/CFT dictionary is still incomplete. It will be very interesting to establish precisely how the novel (A)dS$_4$ higher-spin charges constructed in this work fit into the standard holographic dictionary. One natural guess is that our higher-spin charges are bulk dual to the CFT$_3$ lightray operators associated with the stress tensor recently shown to generate the $\mathcal{L}_{\Lambda} w_{1 + \infty}$ algebra \cite{Strominger:2026yrh}. However, at least naively, the two constructions of the higher-spin $\mathfrak{sl}(2,\mathbb{R})$ primaries differ in both the nature of the defects they correspond to (timelike/spacelike vs. null) and the way in which the higher-spin modes are selected. The construction here parallels that in flat space where the higher-spin modes are picked out by a Mellin transform with respect to the retarded time along the boundary, while in \cite{Strominger:2026yrh} they appear as Fourier modes of light-ray operators along a spacelike circle. Nevertheless, these constructions are tied together by the idea of searching for primary operators with respect to a Lorentz subalgebra of the (A)dS$_4$/CFT$_3$ isometry algebra. It will be interesting to better understand the space of CFT and bulk/spacetime defects with this property.

While the recursion relations \eqref{Lrec} are valid for low-spin for any $\Lambda$, the construction of the  $\Lambda$-deformed charges and their algebra relied on a perturbative expansion in $\Lambda$. It will be interesting to understand if there is an avatar of these symmetries at finite $\Lambda$, or conversely, if these symmetries are a universal, emergent property of holographic CFTs in the large-$N$ limit. This question could be concretely addressed in the context of exact AdS$_4$/CFT$_3$ dualities \cite{Klebanov:2002ja,Bagger:2006sk,Aharony:2008ug,Drukker:2010nc} or perhaps uplifted to other stringy dualities in higher-dimensions provided that the construction of higher spin charges can be generalized. The latter question is of independent interest and will be explored elsewhere. 

Most importantly, it remains an important open problem to establish whether these higher-spin symmetries have any implications on gravitational observables. Having paved the way towards sharply reframing this question in the setting of AdS/CFT, we hope that an answer to this question is now well within reach.


\begin{acknowledgments}
\textit{Acknowledgments.}--- We thank N\'uria Navarro for initial collaboration and Stefano Ansoldi, Laurent Freidel, Ahmed Sheta, Kostas Skenderis, Andrew Strominger and C\'eline Zwikel for discussions. The authors acknowledge the use of ChatGPT for assistance with computations and algebraic checks. The authors are fully responsible for the scientific ideas, analytical strategy, derivations, conclusions and manuscript preparation. M.M. is funded by STFC DTP under grant reference UKRI1780.   
\end{acknowledgments}


\pagebreak \widetext
\section*{End Matter}

\subsection{The $\mathcal{L}_{\Lambda}w_{1 + \infty}$ algebra}

The $\Lambda$-deformed $w_{1 + \infty}$ algebra $(\mathcal{L}_{\Lambda}w_{1 + \infty})$ stated in terms of the modes of the $\mathcal{L}_{\Lambda}w_{1 + \infty}$ generators in \cite{Taylor:2023ajd, Bittleston:2024rqe} is equivalent to the following algebra 
\begin{equation}
\label{eq:charge-aspect-alg}
\begin{split}
    \{ w_s(z), w_{s'}(z')\} &= (s' + 1) \p \delta(z,z') w_{s+s'-1}(z') - (s + 1)\p'\delta(z,z') w_{s+s'-1}(z)\\
    &- \Lambda \left[ (s' - 1) \bar{\p}\delta(z,z') w_{s+s'+1}(z') - (s - 1) \bar{\p}'\delta(z,z')w_{s+s'+1}(z)\right].
    \end{split}
\end{equation}
Following \cite{Freidel:2021ytz}, we expect the $\mathcal{L}_{\Lambda}w_{1 + \infty}$ generators to be identified with spacetime charge aspects carrying the same quantum numbers with respect to an $\mathfrak{so}(3,1)$ subalgebra of the (A)dS$_4$ isometry algebra, namely
\begin{equation}
w_s(z) \sim Q_s(z),
\end{equation}
up to normalization. For the $\Lambda = 0$ algebra, there is always some freedom in this choice since \eqref{eq:charge-aspect-alg} is invariant under the rescaling 
\begin{equation}
\label{eq:redef}
    w_s(z) \rightarrow i^{s-1} w_s(z).
\end{equation}
On the other hand, for $\Lambda \neq 0$ and
under the same rescaling, the two terms on the RHS of \eqref{eq:charge-aspect-alg} acquire a relative sign. This sign may be eliminated by further rescaling
\begin{equation}
    \Lambda \rightarrow -\Lambda.
\end{equation}

In order to obtain a match between the algebra of the charge aspects obtained in this work and that of \cite{Taylor:2023ajd, Bittleston:2024rqe}, a complex phase of the form \eqref{eq:redef} needs to be included in the identification of the bulk charge aspects and the $w_s$ generators, namely
\begin{equation}
\label{eq:phase}
    \cQ_s[\tau] \equiv i^{1-s}\int_S d^2z ~\tau_s(z) w_s(z).
\end{equation}
 In this case \eqref{eq:charge-aspect-alg} becomes
\begin{equation}
\label{eq:cbl}
    \{\cQ_s[\tau], \cQ_{s'}[\tau'] \} = \cQ_{s+ s' - 1}[[\tau,\tau']_0] + \cQ_{s+ s' + 1}[[\tau,\tau']_{\Lambda}],
\end{equation}
where 
\begin{equation}
\label{eq:br}
\begin{split}
    [\tau,\tau']_0 &= -(s' + 1) \tau'_{s'} \p  \tau_s + (s+1) \tau_s \p\tau'_{s'}, \\
    [\tau,\tau']_{\Lambda} &= -\Lambda \left[(s' - 1) \tau_{s'}' \bar{\p} \tau_s + (s-1) \tau_s \bar{\p}\tau'_{s'} \right]. \\
    \end{split}
\end{equation}
Note that $\tau_s$ have helicity $-s$, while $[\tau, \tau']_0$ and $[\tau, \tau']_{\Lambda}$ have helicity $-s-s'+1$ and $-s-s'-1$ respectively. The helicity introduces a grading on the Lie algebra \cite{Norris1997SchoutenNijenhuis}. The general bracket \eqref{Lbra} reduces to the first and second equations of \eqref{eq:br} upon projecting onto its degree $s + s' - 1$ and $s + s' + 1$ components and restricting to the initial data \eqref{eq:in-da}.

\subsection{Solution to the dual recursion relations}
\label{subsec:dual-rr}

In this appendix we provide the ingredients needed to construct the higher-spin charge aspects to quadratic order in the fields and linear order in $\Lambda$. The complete derivations will appear in \cite{DiGiacomoEtAl2026}.
The solution to the dual recursion relations \eqref{puT} with boundary conditions \eqref{eq:in-da} to $\mathcal{O}(\Lambda,C)$ is
\begin{equation}
    T_n =T_n^{(0,0)}+T_n^{(0,1)}+\Lambda T_n^{(1,0)}+\Lambda T_n^{(1,1)},
\end{equation}
where 
\begin{equation}\label{T00}
    \begin{split}
    T^{(0,0)}_n&=\frac{u^{s-n}}{(s-n)!}\partial^{s-n}\tau, \quad -1\leq n<s,\\
        T_{n}^{(1,0)}&= -\frac{u^{s-n+2}}{(s-n+1)!}\bar\partial\partial^{s-n+1}\tau,\ \ 0\leq n\leq s+1,\\
        T_{-1}^{(1,0)}&=-\frac{s+2}{(s+3)!}u^{s+3}\bar\partial \partial^{s+2}\tau
    \end{split}
\end{equation}
and 
\begin{equation}\label{eq:Tquadratic}
\begin{split}
T_n^{(0,1)}
&= -\frac{1}{2}\sum_{k=0}^{s-n-2}(n+k+3)
   \partial^k(\partial_u^{-1})^{k+1}
   \bigl(C T^{(0,0)}_{n+k+2}\bigr),
\quad  -1\leq n\leq s-2,
\\
T^{(1,1)}_n
&= -\sum_{k=0}^{s-n-1}
   \partial^k(\partial_u^{-1})^{k+1}
   \Bigl[
      \bar\partial T^{(0,1)}_{n+k-1}
      +\frac{n+k+3}{2}C T^{(1,0)}_{n+k+2}
      +(\partial_u^{-1}C)\bar\partial T^{(0,0)}_{n+k+1}
      +(n+k+1)\bar\partial(\partial_u^{-1}C)
       T^{(0,0)}_{n+k+1}
   \Bigr], \\
   &\hspace{400pt} -1\leq n\leq s-1.
\end{split}
\end{equation}

Substituting these expressions into \eqref{eq:nc}, we obtain the linearized charge by adding \eqref{eq:l0} and \eqref{eq:l1} and the quadratic charge
\begin{equation}
\label{eq:quadrlc}
    \cQ_{s, {\rm quad}}^{\Lambda}[\tau] = \cQ_s^{(0,1)}[\tau] +  \Lambda \cQ_s^{(1,1)}[\tau]
\end{equation}
where
\begin{equation}
\label{eq:qch}
    \begin{split}
        \cQ_s^{(0,1)}[\tau] &= -\frac{4}{\kappa^2} \int du \int_S Q_{-2} \sum_{k = 0}^s(k+1) \p^k (\p_u^{-1})^k\left(\frac{u^{s - k}\p^{s - k}\tau}{(s - k)!}  C\right) \\
        \cQ_s^{(1,1)}[\tau] & = -\frac{4}{\kappa^2}\int du \int_S Q_{-2}\left[ -2 \p T_{-1}^{(1,1)} + 2 \p_u^{-1} C \bar{\partial} T_{-1}^{(0,0)} - 2 \bar{\partial} \p_u^{-1} C T_{-1}^{(0,0)} + T_0^{(1,0)} C \right].
    \end{split}
\end{equation}
The two middle terms in the last expression arise by making the time-dependence of $\eth$ explicit using \eqref{comm}. 

\subsection{Taylor-Zhu OPE}

Following the same approach as \cite{Guevara:2021abz}, it was shown in \cite{Taylor:2023ajd} that the OPE of conformal primary gravitons
\begin{equation}
    \begin{split}
        \GTZ{\Delta_1}(z_1,\bz_1)\GTZ{\Delta_2}(z_2,\bz_2)
        &=-\frac{\kappa}{2\bar z_{12}}\sum_{n\ge0}
        B(\Delta_1-1+n,\Delta_2-1)\frac{z_{12}^{n+1}}{n!}\p^n\GTZ{\Delta_1+\Delta_2}(z_2,\bz_2)\\
        &+\frac{\kappa\Lambda}{2}\frac{\Delta_1+\Delta_2}{\bar z_{12}^{\,2}}
        \sum_{n\ge0}B(\Delta_1-2+n,\Delta_2-2)\frac{z_{12}^n}{n!}\p^n\GTZ{\Delta_1+\Delta_2-2}(z_2,\bz_2)\\
        &+\frac{\kappa\Lambda}{2}\frac{\Delta_1}{\bar z_{12}}
        \sum_{n\ge0}B(\Delta_1-2+n,\Delta_2-2)\frac{z_{12}^n}{n!}\p^n\db\GTZ{\Delta_1+\Delta_2-2}(z_2,\bz_2)
    \la{OPETZ}
    \end{split}
\end{equation}
implies that the associated conformally soft gravitons (or more precisely, conformal primary descendants thereof) obey the $\mathcal{L}_{\Lambda}w_{1 + \infty}$ algebra. One can check that the $\Lambda = 0$ and $\Lambda \neq 0$ terms are independently $\mathfrak{sl}(2,\mathbb{R})$ covariant, irrespective of the $\Delta$-dependent prefactors multiplying the $\Lambda$ corrected OPE blocks. These factors are nevertheless uniquely fixed by demanding that this OPE implies the $\mathcal{L}_{\Lambda} w_{1 + \infty}$ algebra. The same factors appear naturally in the algebra of the $\Lambda$-corrected linear and quadratic charges \eqref{dLG}.

\bibliography{biblio-w-PRL.bib}

\end{document}